\documentstyle[11pt]{article}
\begin{document}
\setlength{\baselineskip}{15pt}
\title{Separation of variables in the Jacobi identities}
\author{Benito Hern\'{a}ndez--Bermejo $^1$ \and V\'{\i}ctor Fair\'{e}n }
\date{}

\maketitle

\noindent{\em Departamento de F\'{\i}sica Matem\'{a}tica y Fluidos, 
Universidad Nacional de Educaci\'{o}n a Distancia. Senda del Rey S/N, 28040 
Madrid, Spain.}

\mbox{}

\begin{center} 
{\bf Abstract}
\end{center}
\noindent
A new family of $n$-dimensional solutions of the Jacobi identities is 
characterized. Such a family is very general, thus unifying in a common 
framework many different well-known Poisson systems seemingly unrelated. This 
unification is not only conceptual, but also allows the development of 
general global methods of analysis.

\mbox{}

\mbox{}

\mbox{}

\mbox{}


\mbox{}

\mbox{}

\mbox{}

\mbox{}

\mbox{}

\mbox{}

\mbox{}

\mbox{}

\mbox{}

\mbox{}

\mbox{}

\mbox{}

\mbox{}

\mbox{}

\mbox{}

\mbox{}

\mbox{}

\mbox{}

\noindent $^1$ Corresponding author. E-mail: bhernand@apphys.uned.es

\pagebreak
\begin{flushleft}
{\bf 1. Introduction}
\end{flushleft}

Poisson structures \cite{lich1}--\cite{olv1} have an important presence in 
all fields of Mathematical Physics, such as dynamical systems 
theory \cite{nut1}--\cite{byv2}, fluid dynamics \cite{hs1}, 
magnetohydrodynamics \cite{myg1}, optics \cite{dht3}, continuous media 
\cite{dv1}, etc. Describing a given physical system in terms of a Poisson 
structure opens the possibility of obtaining a wide range of information 
which may be in the form of perturbative solutions \cite{cyl1}, invariants 
\cite{byv3,tbym}, nonlinear stability analysis \cite{hyct}, bifurcation 
properties and characterization of chaotic behaviour \cite{dht3}, or 
integrability results \cite{mag1}, to cite a few. 

Mathematically, a finite-dimensional dynamical system is said to have a 
Poisson structure if it can be written in terms of a set of ODEs of the form: 
\begin{equation}
    \label{nham}
    \dot{x}^i = \sum_{j=1}^n J^{ij} \partial _j H \; , \;\:\; i = 1, \ldots , n, 
\end{equation} 
where $H(\mbox{{\bf x}})$, which is usually taken to be a time-independent 
first integral, plays the role of Hamiltonian function. The 
$J^{ij}(\mbox{{\bf x}})$ are the entries of an $n \times n$ matrix ${\cal J}$ 
which may be degenerate in rank ---known as the structure matrix--- and they 
have the property of being solutions of the Jacobi identities: 
\begin{equation}
     \label{jac}
     \sum_{l=1}^n ( J^{li} \partial_l J^{jk} + J^{lj} \partial_l J^{ki} + 
     J^{lk} \partial_l J^{ij} ) = 0 
\end{equation}
In (\ref{jac}), $ \partial_l $ means $ \partial / \partial x^l$ and indices 
$i,j,k$ run from 1 to $n$. The $J^{ij}$ must also verify the additional 
condition of being skew-symmetric:
\begin{equation}
     \label{sksym}
     J^{ij} =  - J^{ji} \;\:\:\:\: \mbox{for all} \:\; i,j
\end{equation}
The possibility of describing a given 
finite-dimensional dynamical system in terms of a Poisson structure is still 
an open problem \cite{nut1,gyn1,hyg1,pla1,byv2,per1,koz1}. The source of 
the difficulty arises not only from the need of a known first integral 
playing the role of the Hamiltonian, but mainly due to the necessity of 
associating a suitable structure matrix to the problem. In other words, 
finding an appropriate solution of the Jacobi identities (\ref{jac}), 
complying also with the additional conditions (\ref{sksym}), is unavoidable. 
This explains, together with the intrinsic mathematical interest of the 
problem, the permanent attention deserved in the literature by the obtainment 
and classification of skew-symmetric solutions of the Jacobi identities. In 
the simplest case of three-dimensional flows it has been possible to rewrite 
equations (\ref{jac}--\ref{sksym}) in more manageable forms allowing the 
determination of some families of solutions \cite{gyn1,hyg1}. 
However, this strategy is not applicable when analyzing the general 
$n$-dimensional problem (\ref{jac}--\ref{sksym}). In such a case, the 
present-day classification of solutions of (\ref{jac}--\ref{sksym}) can be 
summarized, roughly speaking, as a sequence of families of solutions having 
increasing nonlinearity: constant structures (of which the well-known 
symplectic matrix is just a particular case), linear (i.e. Lie-Poisson) 
structures \cite{lie1}, affine-linear structures \cite{bha1} and finally 
quadratic structures \cite{pla1,byv2,byr1,lyx1}. 

In this letter we present a new family of skew-symmetric, separable solutions 
of the Jacobi identities. Due to the unusual generality of such a family 
---it consists of $n$-dimensional solutions not limited to a given degree of 
nonlinearity--- many known Poisson structures appear embraced as particular 
cases, thus unifying many different systems seemingly unrelated. As we shall 
see, this unification has relevant consequences for the analysis of such 
systems, since it leads to the development of a common framework for the 
determination of many important properties such as the symplectic structure 
or the Darboux canonical form. Such properties can now be characterized 
globally in a unified and very economic way.

The structure of the article is as follows: In Section 2 we establish the 
main results concerning the form of the solutions and their properties. In 
Section 3 we illustrate the previous results by means of a number of 
examples, chosen to show the generality of the solutions and how very 
different systems, in principle unrelated, appear now grouped naturally and 
can be analyzed in a unified manner. We conclude in Section 4 with some final 
remarks.

\mbox{}

\begin{flushleft}
{\bf 2. Poisson structures arising from separation of variables in the Jacobi 
identities}
\end{flushleft}

Let $\{ \varphi ^1 (x^1), \varphi ^2 (x^2), \ldots , \varphi ^n (x^n) \}$ 
be a set of nonvanishing $C^1$ functions defined on a subset 
$\Omega \subset I \!\! R^n$. The need for the nonvanishing condition 
$\varphi ^i (x^i) \neq 0$ in $\Omega$ will become clear in what follows. In 
the context of this article we define a separable matrix as an $\, n \times 
n \,$ matrix defined in $\Omega$ which is of the form:
\begin{equation}
     \label{msep}
     S^{ij} =  a^{ij} \varphi ^i(x^i) \varphi ^j(x^j) \;\:,\;\:\;\: 
         a^{ij} \in I \!\! R  \;\:, \;\:\; a^{ij} = -a^{ji} \;\:\;
         \mbox{for all} \;\:\; i , j
\end{equation}
Obviously, every separable matrix thus defined is skew-symmetric. Moreover, 
every separable matrix is also a solution of the Jacobi identities 
(\ref{jac}). Therefore, {\em every separable matrix is a structure matrix.\/} 

To prove this, we only need to substitute matrix $S^{ij}$ (\ref{msep}) 
into the Jacobi equations (\ref{jac}). We first arrive to terms of the form:
\begin{equation}
     \label{dem1}
      S^{li} \partial_l S^{jk} = a^{li}a^{jk} \varphi^{l}\varphi^{i} 
      (\delta^{lj}\dot{\varphi}^{j}\varphi^{k} + 
      \delta^{lk}\varphi^{j}\dot{\varphi}^{k})
\end{equation}
where $\dot{\varphi}^{i}$ means $\mbox{d} \varphi^{i}/ \mbox{d} x^i$ and 
$\delta^{ij}$ is Kronecker's delta. Similar expressions are found for the 
other combinations of indexes. Grouping into (\ref{jac}) and simplifying the 
deltas we arrive at:
\[
     \sum_{l=1}^n ( S^{li} \partial_l S^{jk} + S^{lj} \partial_l S^{ki} + 
     S^{lk} \partial_l S^{ij} ) = 
\]
\begin{equation}
      \varphi^{i}\varphi^{j}\varphi^{k} 
            [ \dot{\varphi}^{j} ( a^{ji} a^{jk} + a^{ij} a^{jk} ) +
              \dot{\varphi}^{k} ( a^{jk} a^{ki} + a^{kj} a^{ki} ) +
              \dot{\varphi}^{i} ( a^{ki} a^{ij} + a^{ik} a^{ij} ) ] = 0
\end{equation}
due to the skew-symmetry of the $a^{ij}$. This demonstrates one of the main 
results of this paper.

Therefore, in what follows we shall analyze the main properties of the 
separable structure matrices. According to the previous result, these are all 
of the form:
\begin{equation}
     \label{jsep}
     J^{ij} =  a^{ij} \varphi ^i(x^i) \varphi ^j(x^j) \;\:,\;\:\;\: 
         a^{ij} \in I \!\! R  \;\:, \;\:\; a^{ij} = -a^{ji} \;\:\;
         \mbox{for all} \;\:\; i , j
\end{equation}
where the $\varphi ^{i}$ are nonvanishing arbitrary $C^1$ functions. 

We shall start by considering the determination of the Casimir invariants. 
This can be done with only the knowledge of the structure matrix, since they 
are Hamiltonian independent. In order to solve this problem, the most 
efficient possibility is to use the abbreviated method developed by
Hern\'{a}ndez--Bermejo and Fair\'{e}n \cite{byv3}. For this, we first notice 
that Rank(${\cal J}$) $=$ Rank($A$) in $\Omega$, where $(A)^{ij}=a^{ij}$. 
This is a consequence of the nonvanishing character of the $\varphi^{i}$ in 
$\Omega$. According to \cite{byv3}, this associates naturally the Casimir 
invariants to the kernel of matrix $A$ and leads to the determination of 
their form: If
\begin{equation}
\label{cas}
    C = \sum _{j=1}^{n} k^j \int \frac{\mbox{d}x^{j}}{\varphi^{j}(x^{j})}
    \;\: , \;\:\;\:\;\: \mbox{{\bf k}} = (k^1,k^2, \ldots, k^n)^T \; \in \; 
    \mbox{Ker}(A)
\end{equation}
where the superscript $^T$ denotes the transpose of a matrix, then $C$ is a 
Casimir function of ${\cal J}$. In addition, we have that 
dim$\{ \mbox{Ker}(A)\} = n - \mbox{Rank}(A) \equiv m$, and thus there are $m$ 
linearly independent vectors that span this kernel. Let 
$\{ \mbox{{\bf k}}_{(1)},\mbox{{\bf k}}_{(2)}, \ldots ,\mbox{{\bf k}}_{(m)}\}$ 
be a basis of Ker($A$), and let 
\begin{equation}
    C_{(i)} = \sum _{j=1}^{n} k_{(i)}^j \int 
    \frac{\mbox{d}x^{j}}{\varphi^{j}(x^{j})}
    \;\: , \;\:\;\:\;\: i = 1,2, \ldots, m
\end{equation}
be the $m$ Casimir invariants associated to this basis. Then, if we evaluate 
the Jacobian matrix associated with $\{ C_{(1)}, C_{(2)}, \ldots ,C_{(m)} \}$
it can be immediately seen that such Casimir invariants are functionally  
independent. Since $m$ is also the corank of ${\cal J}$, this demonstrates 
that $\{ C_{(1)}, C_{(2)}, \ldots , C_{(m)} \}$ is a complete set of 
independent Casimir functions of the structure matrix (\ref{jsep}), any other 
Casimir invariant being a functional combination of them. Therefore, the 
symplectic foliation of the separable structure matrices can be completely 
determined from the kernel of the constant matrix $A$, which is a significant 
simplification of the problem.

We now examine the reduction to the Darboux canonical form. For this, we 
shall take into account the equation for the transformation of the structure 
matrix under general diffeomorphisms, $y^i = y^i(x)$:
\begin{equation}
\label{trnsj}
    \tilde{J}^{ij}(y) = \sum_{k,l=1}^n \frac{\partial y^i}{\partial x^k}
    J^{kl}(x) \frac{\partial y^j}{\partial x^l}
\end{equation}
We thus introduce the following diffeomorphic transformation, which is 
globally defined in $\Omega$:
\begin{equation}
\label{tdar1}
   y^i = \int \frac{\mbox{d}x^i}{\varphi^{i}(x^i)} \:\; , \:\;\:\;\:\; 
   i = 1, \ldots , n
\end{equation}
When (\ref{jsep}) and (\ref{tdar1}) are substituted in (\ref{trnsj}), we 
obtain:
\begin{equation}
   \tilde{J}^{ij}(y) = a^{ij} \:\; , \:\;\:\;\:\; \mbox{for all} \;\: i , j
\end{equation}
In other words, we have transformed the matrix in such a way that now 
$\tilde{{\cal J}} = A$ is a structure matrix of constant entries. In 
addition to this, we now perform a second transformation, which is also 
globally defined:
\begin{equation}
\label{trp}
   z^i = \sum _{j=1}^n P^{ij} y^j \:\; , \:\;\:\;\:\; i = 1, \ldots , n
\end{equation}
where $P$ is a constant, $n \times n$ invertible matrix. According to 
(\ref{trnsj}), the structure matrix $\tilde{J}$ now becomes:
\begin{equation}
   \label{trp2}
   \hat{\cal J} = P \cdot \tilde{\cal J} \cdot P^T = P \cdot A \cdot P^T
\end{equation}
It is well known \cite{ayr1} that matrix $P$ in (\ref{trp2}) can be chosen to 
have:
\begin{equation}
\label{jham}
   \hat{\cal J} = \mbox{diag} (D_1, D_2, \ldots , D_{r/2}, 
                  \overbrace{ 0, \ldots ,0 }^{(n-r)})
\end{equation}
where $r = \mbox{Rank}(A)$ is an even number because $A$ is skew-symmetric, 
and 
\begin{equation}
\label{jham2}
   D_1 = D_2 = \ldots = D_{r/2} = 
   \left( \begin{array}{cc} 0 & 1 \\ -1 & 0 \end{array} \right)
\end{equation}
With (\ref{jham}--\ref{jham2}), the structure matrix has been reduced to the 
Darboux form, since $\hat{\cal J}$ is a direct sum of $2 \times 2$ symplectic 
matrices plus $\, (n-r) \,$ null $1 \times 1$ matrices corresponding to the 
Casimir invariants, which in the Darboux representation are decoupled and 
correspond to the variables $\{ z_{r+1}, \ldots , z_{n} \}$. It is worth 
emphasizing that the reduction has been completed explicitly and globally on 
the domain of interest. This is interesting, since the number of Poisson 
structures for which this can be done is exceedingly limited. Well on the 
contrary, in the present case this is possible in a quite natural way. 

In addition to these advantageous manipulation properties, the separable 
structure matrices embrace and unify many different Poisson structures common 
in the literature. We shall now see a sample in the following section. 

\mbox{}

\begin{flushleft}
{\bf 3. Examples}
\end{flushleft}

\noindent{\em (I) Lotka-Volterra and Generalized Lotka-Volterra systems}

The following kind of separable structure matrices
\begin{equation}
\label{jpla}
   J^{ij} = a^{ij} x^i x^j \;\:,\;\:\;\: a^{ij} \in I \!\! R  \;\:, \;\:\; 
            a^{ij} = -a^{ji} \;\:\; \mbox{for all} \;\:\; i , j
\end{equation}
were first recognized by Plank \cite{pla1} in the characterization of Poisson 
structures of the Lotka-Volterra equations, and were important later in the 
wider case of generalized Lotka-Volterra Poisson systems \cite{byv2}. 
However, particular cases of (\ref{jpla}) had been previously found in 
different contexts, such as plasma physics \cite{pyj1} or population dynamics 
\cite{nut1,gyn1,cyf1} (see also the relativistic Toda equations in the next 
example). 

In (\ref{jpla}) we have $\varphi ^i (x^i) = x^i$, and therefore the Casimir 
invariants are immediately found to be, according to (\ref{cas}), of the 
form:
\begin{equation}
\label{caslv}    
    C = \sum _{j=1}^{n} k^j \ln (x^j) \;\: , \;\:\;\:\;\: \mbox{{\bf k}} = 
    (k^1,k^2, \ldots, k^n)^T \; \in \; \mbox{Ker}(A)
\end{equation}
In the specific case of Lotka-Volterra equations, the first integrals 
(\ref{caslv}) were already noticed by Volterra himself \cite{volt1}, but 
they were not generically recognized as Casimir invariants until Plank's 
work \cite{pla1}. Being Hamiltonian independent, they also appear in more 
general types of models sharing the structure matrix (\ref{jpla}), such as 
those treated in \cite{byv2}.

The first transformation (\ref{tdar1}) necessary to achieve the Darboux 
canonical form now is:
\begin{equation}
\label{tlog}   
   y^i = \int \frac{\mbox{d}x^i}{\varphi^{i}(x^i)} = \ln (x^i) 
   \:\; , \:\;\:\;\:\; i = 1, \ldots , n
\end{equation}
The change of variables (\ref{tlog}) is to be followed by the linear 
transformation (\ref{trp}). This kind of two-step reduction to a classical 
Hamiltonian formulation has been known for long ---outside the framework of 
Poisson structures--- in the particular case of conservative, 
even-dimensional Lotka-Volterra systems \cite{ker1}. The realization that 
such reduction is, in fact, Hamiltonian independent and inherent to structure 
matrices of the kind (\ref{jpla}) was formalized recently in \cite{byv2}.

\mbox{}

\noindent{\em (II) Toda lattice and relativistic Toda lattice}

Toda system, when expressed in Flaschka's variables \cite{dam1} $(\alpha^1, 
\ldots , \alpha^{N-1}, \beta^1, \ldots , \beta^N)$ is a Poisson system with 
brackets $\{ \alpha^i , \beta^i \} = - \alpha^i$, $\{ \alpha^i , \beta^{i+1} 
\} = \alpha^i$, while the rest of the elementary brackets vanish. This 
Poisson bracket corresponds to a separable structure given by: 
\begin{equation}
\label{tlf}
   \begin{array}{rcll}
   \varphi^i(\alpha^i) & = & \alpha^i \; , & i =1, \ldots , N-1 \\
   \varphi^j(\beta^j) & = & 1   \; , & j =1, \ldots , N
   \end{array}
\end{equation}
\begin{equation}
\label{tla}
   A = \left( \begin{array}{cc}
      O_{(N-1) \times (N-1)} & M_{(N-1) \times N} \\
      - M^T_{N \times (N-1)} & O_{N \times N}
      \end{array} \right)
\end{equation}
where the subindexes of the submatrices indicate their sizes, $O$ denotes 
the null matrix and 
\begin{equation}
\label{mm}
   M_{(N-1) \times N} = 
       \left( \begin{array}{cccccc}
         -1  &  1  &  0  &  \ldots &  0  &  0  \\
          0  & -1  &  1  &  \ldots &  0  &  0  \\
          \vdots & \vdots & \vdots & \mbox{} & \vdots & \vdots  \\
          0  &  0  &  0  &  \ldots & -1  &  1
       \end{array} \right)
\end{equation}
It is immediate that the kernel of $A$ is one-dimensional, a basis 
of which is provided by the vector $\mbox{{\bf k}} = ( \overbrace{ \; 0 \; , 
\; \ldots \; , \; 0 \; }^{N-1} , \overbrace{ \; 1 \; , \; \ldots \; , \; 
1 \; }^{N} )^T$. Consequently, from (\ref{cas}) there is only one independent 
Casimir invariant, $C = \sum _{j=1}^{N} \int \mbox{d} \beta^j = 
\sum _{j=1}^{N} \beta^j$, which is the result found in \cite{dam1}. The 
reduction to the Darboux form also becomes straightforward, since we have to 
perform transformation 
(\ref{tdar1})
\begin{equation}
   \begin{array}{rcll}
   \tilde{\alpha}^i & = & \ln (\alpha^i) \; ,  & i =1, \ldots , N-1 \\ 
   \tilde{\beta}^j  & = & \beta^j   \; ,       & j =1, \ldots , N 
   \end{array}
\end{equation}
and then carry out the linear change of variables (\ref{trp}).

Analogously, we consider now the relativistic Toda equations expressed in 
similar variables \cite{dam1} $(\alpha^1, \ldots ,\alpha^{N-1}, \beta^1, 
\ldots , \beta^N)$. Again, it is a Poisson system with brackets $\{ \alpha^i, 
\alpha^{i+1}\} = \alpha^i \alpha^{i+1}$, $\{ \alpha^i,\beta^i\} = -\alpha^i 
\beta^i $, $\{ \alpha^i,\beta^{i+1}\} = \alpha^i \beta^{i+1}$, while the rest 
of the elementary brackets vanish. This Poisson bracket corresponds to a 
separable structure of the form (\ref{jpla}) examined in Example I. 
Therefore, all the considerations made there hold in this context. 

\mbox{}

\noindent{\em (III) Further examples: Constant structure matrices, 
Kermack-McKendric model, circle maps, $2 \times 2$ games}

We end the present section with a brief enumeration of other examples which 
have also deserved some interest in the literature. We shall not elaborate on 
them with the detail of the previous instances, but only outline the most 
interesting features. 

The first example will be that of the constant skew-symmetric structure 
matrices, which appear frequently in very diverse developments. In this 
case we would like to recall that $\varphi ^i (x^i) = 1$, and thus the 
Casimir functions (\ref{cas}) are linear. Notice also that transformation 
(\ref{tdar1}) reduces to the identity, and therefore the reduction to 
classical Hamiltonian form only involves a linear transformation 
(\ref{trp}--\ref{jham}). These are well-known facts that now appear as a 
simple particular case. 

As a second example we touch upon the Kermack-McKendric model 
\cite{nut2,gyn1}, which admits a Poisson structure in terms of matrix:
\begin{equation}
   {\cal J} = \left( \begin{array}{ccc}
          0        &  -rx^1x^2  &  0       \\
          rx^1x^2  &     0      &  -ax^2   \\
          0        &    ax^2    &  0   
       \end{array} \right)
\end{equation}
where the $x^i$ denote the system variables and $\{a,r\}$ are real constants. 
We again have a separable matrix with $\{\varphi ^1 (x^1), \varphi ^2 (x^2), 
\varphi ^3 (x^3) \} = \{ x^1,x^2,1 \}$. Therefore, this example turns out to 
be very similar to the nonrelativistic Toda lattice examined before, as it 
can be seen from (\ref{tlf}--\ref{tla}). We thus find that seemingly 
unrelated problems can be analyzed in a similar, unifying framework.

Next we shall mention the Poisson structures appearing in the study of 
certain circle maps \cite{gyn1}, in which we have:
\begin{equation}
   {\cal J} = \left( \begin{array}{ccc}
                 0        &        0        &  -(x^1)^2(x^3)^2   \\
                 0        &        0        &  -(x^2)^2(x^3)^2   \\
          (x^1)^2(x^3)^2  & (x^2)^2(x^3)^2  &  0   
       \end{array} \right)
\end{equation}
We thus have $\varphi ^i (x^i) = (x^i)^2$, a more nonlinear kind of function. 
The evaluation of the Casimir invariants and the Darboux canonical form do 
not present any special difficulty in this case and are omitted.

Finally we shall consider a very different kind of structure matrix found in 
the context of $2 \times 2$ games \cite{hof1}, in which: 
\begin{equation}
   {\cal J} = \left( \begin{array}{cc}
                   0            &  x^1(1-x^1)x^2(1-x^2)  \\
         -x^1(1-x^1)x^2(1-x^2)  &          0 
       \end{array} \right)
\end{equation}
Now we have that $\Omega = \mbox{int} \{ S_1 \times S_1 \}$ is the interior 
of the product of two probability simplices, and $\varphi ^i (x^i) = 
x^i(1-x^i)$. Obviously there are no Casimir functions in this case. Because 
now $A$ is the $2 \times 2$ symplectic matrix, no linear transformation 
(\ref{trp}) is required and the reduction to the classical Hamiltonian form 
only involves transformation (\ref{tdar1}):
\begin{equation}
   y^i = \int \frac{\mbox{d}x^i}{x^i(1-x^i)} = \ln 
   \left( \frac{x^i}{1-x^i} \right) \;\:,\;\;\:\;\: i = 1,2
\end{equation}
Notice that all the manipulations are properly defined because functions 
$\varphi ^i (x^i)$ are nonvanishing in the domain $\Omega$.

\mbox{}

\pagebreak
\begin{flushleft}
{\bf 4. Final remarks}
\end{flushleft}

In this letter we have presented a new family of skew-symmetric solutions of 
the Jacobi identities which exhibits a number of interesting properties: 
\begin{itemize}
\item It can be found by applying the classical method of separation of 
variables. It is mathematically remarkable that such a technique, in 
principle restricted to the domain of certain linear PDEs, can be applied in 
this context. 

\item The resulting solutions are valid for the $n$-dimensional problem 
(\ref{jac}--\ref{sksym}), independently of the value of $n$, and are not 
limited to a given degree of nonlinearity. Well on the contrary, they can 
contain polynomials of arbitrary degree and general functions as well. 

\item They generalize already known solutions. In fact, they unify in a 
common framework many different Poisson structures and well-known systems 
which seemed to be unrelated, and now appear as particular cases of a common 
family. We have provided a list of them in the examples. 

\noindent And last but not least:

\item This unification is not only conceptual, but allows the development of 
general methods of analysis simultaneously valid for every particular case. 
Specifically, it is possible to determine in an algorithmic and explicit way 
the Casimir invariants and the Darboux canonical form. Moreover, these 
results hold globally in the phase-space ---in contrast with the usual scope 
of Darboux' theorem \cite{olv1}, which only guarantees a local reduction.
\end{itemize}

As we have seen, the direct search of new Poisson structures as solutions of 
the Jacobi identities leads to a unifying perspective of very diverse, and 
seemingly unrelated, systems. This is useful not only from a classification 
point of view ---from which we have obtained a remarkably general family of 
solutions--- but also because it is possible, as we have demonstrated, to 
develop a unified approach for their analysis which becomes much more 
economic and elegant than a heuristic, case-by-case strategy. Obviously, it 
is feasible to establish such techniques with regard to all Hamiltonian 
independent properties, such as those treated above. 

For these reasons, it is the authors' impression that this kind of analysis 
of the Jacobi equations can produce interesting results for further 
generalization and unification of finite-dimensional Poisson structures. In 
particular, two clear lines of investigation are an additional generalization 
of the separable solutions (\ref{jsep}) found in this paper, as well as the 
search of completely different families of structure matrices. These 
possibilities will be the subject of future research. 

\mbox{}

\mbox{}

\begin{flushleft}
{\bf Acknowledgements}
\end{flushleft}

\noindent The authors wish to acknowledge support from the European Union 
(Esprit WG 24490).

\pagebreak

\end{document}